\documentclass[twocolumn,showpacs,preprintnumbers,amsmath,amssymb,nofootinbib]{revtex4}
\usepackage{graphicx,epsfig}
\usepackage{dcolumn}
\usepackage{bm}
\usepackage{multirow}
\usepackage{natbib}
\usepackage{appendix}

\begin{document}

\title{Reconciling MOND and dark matter?}

\author{Jean-Philippe Bruneton$^{1}$\footnote{bruneton@sissa.it}, Stefano Liberati$^{1}$\footnote{liberati@sissa.it}, Lorenzo Sindoni$^{1}$\footnote{sindoni@sissa.it}, Benoit Famaey$^{2}$\footnote{benoit.famaey@ulb.ac.be}}

\affiliation{$^{1}$SISSA/ISAS, Via Beirut 2--4, 34014, Trieste and INFN Sezione di Trieste, Via Valerio, 2, 34127 Trieste, Italy\\$^{2}$Institut d'Astronomie et d'Astrophysique, Universit\'e Libre de Bruxelles, CP 226, Bruxelles, Belgium}
\date{\today}

\begin{abstract}
Observations of galaxies suggest a one-to-one analytic relation between the inferred gravity of dark matter at any radius and the enclosed baryonic mass, a relation summarized by Milgrom's law of modified Newtonian dynamics (MOND). However, present-day covariant versions of MOND usually require some additional fields contributing to the geometry, as well as an additional hot dark matter component to explain cluster dynamics and cosmology. Here, we envisage a slightly more mundane explanation, suggesting that dark matter does exist but is the source of MOND-like phenomenology in galaxies. We assume a canonical action for dark matter, but also add an interaction term between baryonic matter, gravity, and dark matter, such that standard matter effectively obeys the MOND field equation in galaxies. We show that even the simplest realization of the framework leads to a model which reproduces some phenomenological predictions of cold dark matter (CDM) and MOND at those scales where these are most successful. We also devise a more general form of the interaction term, introducing the medium density as a new order parameter. This allows for new physical effects which should be amenable to observational tests in the near future. Hence, this very general framework, which can be furthermore related to a generalized scalar-tensor theory, opens the way to a possible unification of the successes of CDM and MOND at different scales.
\end{abstract}


\maketitle
  
\section{Introduction}

The mass discrepancy puzzle gets definitely harder with accumulation of new data and facts. There are now clear evidences that massive and weakly interacting particles, or Cold Dark Matter (CDM), do not account for all the features of the mass discrepancy problem. Although the standard cosmological model, ($\Lambda$)CDM, has proven to be successful from cosmological \cite{wmap5} down to galaxy cluster scales\footnote{See however \cite{Broadhurst:2008ng} for a different point of view.}, it faces severe problems in explaining the observed phenomenology at galactic scales. 

Often debated issues for the CDM paradigm are the predicted overabundance of satellites \cite{Kauffmann:1993gv,Moore,Kravtsov:2004cm,Kuhlen:2008qj}, the generic formation of CDM cusps \cite{Navarro:1995iw,Diemand} and the related poor quality of fits of spiral galaxies's rotation curves within CDM haloes \cite{McGaugh:2006vv,Gentile:2004tb,Gentile:2006hv,Englmaier}, the problems to form large enough baryonic disks due to their predicted low angular momentum within simulations \cite{Combes,NavStein}, and the departures from the CDM scenario recently found in tidal darf galaxies \cite{Bournaud:2007sz, Gentile:2007gp,Milgrom:2007jp}. 

It is commonly thought that these problems may be due to the fact that physics of baryons is not incorporated rigorously enough in cosmological simulations down to the galactic scales. However, even taking into account these complicated and history dependent feedbacks would hardly explain another general issue: the observed {\it universal} conspiracy between the baryonic and DM distributions in spiral galaxies, {\it independent} of the particular history of each stellar system. A well-known signature of this conspiracy is that the ``asymptotic" velocity to the fourth in spirals is proportional to the total baryonic mass of the galaxy, a law that is known as the baryonic Tully-Fisher relation \cite{Tully:1977fu, McGaugh:2000sr}. In a DM picture, this requires a fine balance between the total amount of DM and the total amount of baryons in galaxies ranging over five decades in mass. Moreover, the Tully-Fisher relation can be extended to a much more constraining law, namely that there exists a universal one-to-one correspondence between the integrated baryonic mass and the integrated dark mass \textit{at all radii} in rotating galaxies, and not only in the outskirts \cite{McGaugh:2005er,Famaey:2006iq,Sanders:2008iy}. This relation, known as Milgrom's law\footnote{Note that such a relation may also be interpreted differently, for instance as a ``radial Tully-Fisher" relation \cite{Yegorova:2006wv} or through the ``Universal Rotation Curve" phenomenology, see \cite{Persic,Gentileurc}.} \cite{Milgrom:1983ca}, involves an acceleration scale $a_0 \sim 10^{-10}m/s^{2}$, below which dark matter starts to dominate over baryonic matter in galaxies, and whose origin is far from clear. 

It has been suggested that CDM could naturally start to dominate in all galaxies beyond a transition radius where the acceleration is close to $a_0$ \cite{KapTurn}, but this interpretation of Milgrom's law within the CDM paradigm failed to reproduce the full phenomenology encapsulated in the law \cite{Milgromkt}. Alternatively, one could consider the possibility that, in a dark matter picture, there exists a mechanism at galactic scales that eventually correlates baryonic matter to its dark counterpart in an universal way. This is very unlikely without specific interactions or specific properties of dark matter. Indeed, assuming that this tight balance between baryons and dark particles arises from the formation process conflicts with the fact that CDM has to be almost non-interacting, and hence subject to very different influences than those affecting baryons through the ages whereas each galaxy develops its own history of instabilities, feedbacks, merging, etc. \cite{Milgrom:2007cs}. This naturally leads to consider seriously the idea that dark matter, if it exists, may have more complicated interactions with baryons than was initially thought. Identifying these interactions first requires a better understanding of the aforementioned conspiracy, and this is provided by a competing paradigm, namely Modified Newtonian Dynamics, or MOND \cite{Milgrom:1983ca,Milgrom:2008rv}. 

MOND's starting point is a departure from standard dynamics below the aforementioned scale $a_0$. Equivalently, in the non-relativistic and weak-field approximation, the Newtonian gravitational potential $\Phi$ is assumed to satisfy a modified Poisson equation~\cite{Milgrom:1983ca,BM84}:
\begin{equation}
\label{Mond}
\nabla \cdot \left( \mu \left( \frac{\left|\nabla \Phi\right|}{a_0}\right) \nabla \Phi\right) = 4 \pi G \rho,
\end{equation}
where $\mu(x)$ is a function that differs from $1$ only for small $x$ (for which it tends toward $\mu=x$), and where $\rho$ is the baryonic matter density. Surprisingly enough, this simple rule is able to account for rotation curves of most spiral galaxies and to naturally explain the above mentioned baryon-dark matter conspiracy. Moreover, MOND also explains naturally many observed features of galactic dynamics \cite{Sanders:2002pf, Sanders:2008iy, Milgrom:2007cs,Milgrom:2008rv}. Although the above formulation of MOND is not relativistic, there are by now well-defined relativistic field theories of MOND \cite{Bekenstein:2004ne,Zlosnik,Halle} (see \cite{Bruneton:2007si} for a review). Note that such relativistic theories are generally made easily compatible with solar system tests, given that accelerations are typically larger than $a_0$ in this environment.

The interesting point is that, just as the $\Lambda$CDM model brilliantly accounts for observational cosmology and for cluster dynamics but fails at galactic scales, MOND explains quite successfully galactic dynamics, but does not account for all the mass discrepancy at cluster scales (e.g. \cite{Sanders:2002ue, Pointecouteau:2005mr, Milgrom:2007cs, Angus:2007mn, Clowe:2003tk}), and has difficulties in reproducing the angular power spectrum of the Cosmic Microwave Background (CMB)~\cite{Skordis:2005xk}. 

A straightforward solution is to introduce an additional dark sector also in such a MONDian Universe. It was shown that ordinary neutrinos, if massive enough ($m \sim 2 eV$, i.e. the experimental upper limit), can indeed account for the mass discrepancy in many clusters \cite{Sanders:2007zn, Angus:2006ev} without affecting much galactic dynamics because they cannot collapse at this scale, except for galaxies located at the center of clusters (see e.g. \cite{Richtler:2007jd}) that indeed lack some dark mass even in a MONDian theory. Such a Hot Dark Matter (HDM) also helps in reproducing the CMB data in a MONDian Universe. 

However, a recent work showed that a remaining class of clusters would still be unaccounted for with massive ordinary neutrinos \cite{Angus:2007mn}, meaning that the HDM should be made of slightly more massive neutral fermions, like sterile neutrinos predicted by some extensions of the standard model. It has even been argued that this kind of light particles could solve all the problems of MOND \cite{Angus:2007mn,Angus:2008qz}, although many quantitative tests remain to be performed. 

Furthermore, one may wonder whether modifying gravity (usually through the addition of a dynamical scalar and/or vector field  \cite{Bekenstein:2004ne,Zlosnik,Halle}), and moreover adding an unknown particle is not a price too high to pay in order to account for the mass discrepancy at all scales. It appears that, given the successes of CDM on large scales and those of MOND at galactic scales, another route is actually possible: one might keep the collisionless DM phenomenology on large scales, and try to introduce MONDian phenomenology at galactic scales by attributing to dark matter some non-negligible interactions with standard matter and/or gravity. In that case, one would thus start with a dark sector which would itself be the source of Milgrom's law in galaxies (see also  \cite{Blanchet:2006yt,Blanchet:2008fj, Zhao} for alternative ideas in the same directions). 

More precisely, then, the DM-baryon conspiracy, i.e. the MOND regime, should arise only as a consequence of the particular coupling of DM with baryons. We thus hereafter couple standard matter and dark matter to the metric in the same way (even though doing otherwise could be a promising approach to explain dark energy, see \cite{Alimi}), and look for the possible coupling between dark matter, standard matter and gravity that could lead to the MOND phenomenology in galaxies, while keeping a collisionless DM phenomenology at larger scales. This could open the way to a unification of both the successes of CDM and MOND at different scales.  

We proceed to possible models in Sect.~\ref{Section2}, and discuss their specific features and predictions in Sect.~\ref{Section3}. Conclusions and perspectives are drawn in Sect.~\ref{Section4}.

\section{Framework}
\label{Section2}
We work within an effective theory which is mainly aimed at catching the essential features of a possible DM-baryon coupling creating MONDian phenomenology in galaxies. As we are mainly concerned with this new and non-negligible interaction between dark and standard matter, we do not bother considering the weak interaction between the baryonic and dark sectors which is predicted by physics beyond the standard model. As a consequence, we will not deal with any specific particle physics model for the dark sector, and simply assume it to be well described by a simple scalar field (fluid approximation).

\subsection{Actions}
Our framework is thus defined by a very general action of the following form: 
\begin{equation}
\label{Action1}
S = S_{\textrm{EH}}[g_{\mu\nu}] + S_{\textrm{SM}}[\psi, g_{\mu\nu}]+S_{\textrm{DM}}[\chi, g_{\mu\nu}]+S_{\textrm{Int}}[\chi,\psi,g_{\mu\nu}],
\end{equation}
where $S_{\textrm{EH}}$ stands for the standard Einstein-Hilbert action with a cosmological constant term (using sign conventions of \cite{MTW}), $\psi$ is collectively denoting the Standard Model fields, $\chi$ represents the DM field, and $S_{\textrm{Int}}$ is some interaction term between dark matter, gravity, and standard matter. We chose a canonical action for the scalar field $\chi$
\begin{equation}
\label{ActionDM}
S_{\textrm{DM}}[\chi, g_{\mu\nu}]= -\frac{c^3}{8 \pi G} \int \sqrt{-g} d^4x \left\{ \left(\partial \chi \right)^2 + \frac{m^2 \chi^2}{2} \right\},
\end{equation}
so that the theory reduces to $\Lambda$CDM when the interaction between DM and standard matter is negligible, and if a high value of the mass $m$ is chosen (see Sect.~\ref{MoCDM}). Let us stress again, that here the fields $\psi$ and $\chi$ are summarizing the matter fields present, in a sort of fluid approximation which traces over the microscopical properties of the fundamental fields. While this operation necessarily makes us blind towards the microphysics of the matter fields, it is certainly a legitimate step to be taken when considering macroscopic configurations like galaxies or clusters of galaxies. In this sense, we will assume that the fluid approximation is holding within the scales considered in our analysis.

One can in principle assume many different forms for the interaction term in Eq.(\ref{Action1}). However, in order to be compatible with MONDian phenomenology, it should produce in some regime an extra-force that applies to baryonic objects regardless of their internal structure or movement. In this regime the net effect of the interaction term is therefore an effective metric along which standard matter propagates:
\begin{equation}
\label{MetriqueEffective}
S_{\textrm{SM}}[\psi, g_{\mu\nu}] + S_{\textrm{Int}}[\chi,\psi,g_{\mu\nu}] \approx S_{\textrm{SM}}[\psi, g_{\mu\nu}+h_{\mu\nu}].
\end{equation}
As a consequence, we can identify up to order $\mathcal{O}(h^2)$ the required interaction term as:
\begin{equation}
\label{SInteraction}
S_{\textrm{Int}}[\chi,\psi,g_{\mu\nu}]= \frac{1}{2 c} \int \sqrt{-g} d^4x \, T_{\textrm{matter}}^{\mu\nu} h_{\mu\nu},
\end{equation}
where $T_{\textrm{matter}}^{\mu\nu} (\psi, g)$ is the stress-energy tensor of matter defined with respect to the metric $g_{\mu\nu}$, and $h_{\mu\nu}(\chi, \psi, g)$ is some rank-two symmetric tensor whose structure will be discussed in Sect.~\ref{Section2B}.

Noticeably, this class of theories is a natural extension of the so called scalar-tensor theories. Indeed, when written in the Einstein frame, these theories can be cast in the form \eqref{Action1},\eqref{ActionDM},\eqref{MetriqueEffective},\eqref{SInteraction} by identifying $h_{\mu\nu}= \left(A^2(\chi) -1\right) g_{\mu\nu}$, where $A^2(\chi)$ is the usual conformal factor \cite{Damour:1992we}. Here however, $h_{\mu\nu}$ can assume a more complicated form, through, for instance, dependence in derivatives of $\chi$ or matter fields themselves (see Sect.~\ref{Section2B}). This formalism is also able to describe, as a special case, the so-called disformal coupling introduced in \cite{Bek92}. In contrast with early expectations~\cite{Bekenstein:1993fs}, such a non-minimal scalar-tensor formalism \cite{Bruneton:2007si} is able to enhance gravitational lensing as well as rotation curves, exactly in the same way as in \cite{Bekenstein:2004ne}.

We can also use the lessons of scalar-tensor theories of gravity about the Einstein and Jordan frames description to get a different viewpoint on this model. We saw that the theory is bimetric up to order $\mathcal{O}(h^2)$, by construction. This enables us to see $\chi$ as a new degree of freedom for the gravitational field (up to this this order), in the sense that the force produced by $\chi$ acts universally on standard matter. This however suggests to rewrite the full action in terms of the metric $\tilde{g}=g+h$. If we neglect higher order terms in $h^{2}$, the action  \eqref{Action1} now reads:
\begin{eqnarray}
S= S_{EH}[\tilde{g}] + S_{SM}[\tilde{g},\psi] + S_{DM}[\tilde{g},\tilde{\chi}] \\ \nonumber
- \frac{1}{16\pi G} \int d^{4}x \sqrt{-\tilde{g}} \left( G^{\mu\nu}(\tilde{g})+\Lambda\tilde{g}^{\mu\nu}\right)
h_{\mu\nu} ,
\end{eqnarray}
where we have neglected a total derivative which does not affect the equations of motion, and we have implicitly transformed the DM field $\chi \to \tilde{\chi}$. Here $G^{\mu\nu}(\tilde{g})$ is the Einstein tensor built with $\tilde{g}$. 
So, in this equivalent representation of the theory, we see that the the dark matter field has to be non-minimally coupled to the metric $\tilde{g}$ through $h_{\mu\nu}(\chi,g)$. However, it is also possible to accommodate in the model the case in which {\em all} the matter fields are non-minimally coupled by assuming $h_{\mu\nu}=h_{\mu\nu}(\chi, \psi, g)$.
 
The non-minimal coupling to geometry is not totally unexpected. Indeed, while the usual minimal coupling is actually motivated only by simplicity arguments, the non-minimal coupling between matter fields and gravity is required for many consistency reasons. For instance, it is imposed by the Einstein equivalence principle \cite{Sonego:1993fw}, or also by renormalizability of quantum field theories in curved spacetimes \cite{Buchbinder:1992rb}. At the same time, it is precisely this non-minimal coupling that allows an interacting DM model to produce a MONDian regime. MOND will thus follow from such a non-trivial gravitational coupling between the matter fields, which, again, is an additional ingredient {\it beyond} the non-geometric interactions between the dark matter and baryonic fields predicted in extensions of the Standard model. That MONDian phenomenology is indeed recovered of course depends on the precise structure of $h_{\mu\nu}$, which we are now going to discuss. 
  
\subsection{Defining $h_{\mu\nu}$}
\label{Section2B}

Having written in Eq.(\ref{ActionDM}) a canonical term for the field $\chi$, it is clear that the existence of a MONDian regime needs the fundamental scale $a_0$ to appear in the interaction term. To this end we can make use of the work reported in \cite{Bruneton:2007si}, where it was shown that both the non-quadratic Lagrangian for the scalar field and the effect of the vector field in TeVeS-like theories can be mimicked by the use of derivatives of $\chi$ alone inside $h_{\mu\nu}$ (sections III.C and V.B of \cite{Bruneton:2007si}). Therefore, the very basics of a relativistic theory of MOND, meaning the reproduction of the flattening of rotation curves of spirals and of the abnormal gravitational lensing, can be recovered in such a model.

In \cite{Bruneton:2007si}, a MONDian metric $g+h$ was constructed using both the values of $\chi$ and its derivatives in the case of negligible mass for the field $\chi$ (see  Eq.(5.11) in \cite{Bruneton:2007si}). However, in our case the $\chi$ field mass is not a priori negligible and hence the above construction must be generalized. The explicit form of $h$ is derived in the appendix and can be cast in the form:
\begin{equation}
h_{\mu\nu}= h_{\mu\nu}^{\textrm{MOND}} \equiv A(\chi,\partial \chi) g_{\mu\nu} + B(\chi,\partial \chi) \partial_{\mu} \chi \partial_{\nu} \chi,
\label{hMond}
\end{equation} 
where  
\begin{eqnarray}
\label{AandB}
A&=& \left(e^{\alpha \chi} - \frac{\chi}{\alpha}  X e^{m \varpi} u(X)\right)^2 - 1,\nonumber\\
B&=& -4 \frac{\chi}{\alpha} \frac{X e^{m \varpi}}{(\partial \chi)^2}.
\end{eqnarray}
Here $X$, $\varpi$ and the function $u$ are defined in the appendix, and $\alpha$ is a pure number characterizing the coupling of $\chi$ to matter. It is worth noting that the field equation for $\chi$ derived from action \eqref{Action1} is then highly non-linear in low acceleration environments (except in vacuum where $T_{\mu\nu}^{\textrm{mat}}=0$). This prevents the possibility to univocally define single particle states. Given that pure MONDian effects arise when the interaction term can not be treated perturbatively, it follows that in MOND-behaving regions one should not consider space as filled by free DM particles.  On the contrary, when one deals with environments where the coupling term is negligible, i.e. where the theory indeed reduces to $\Lambda$CDM to the leading order, a particle like description of dark matter is indeed appropriate. 
From the equations in the appendix, it appears that the relevant order parameter determining the above transition in static systems is truly the gravitational acceleration, just as in standard MOND. 

Of course, this is the most simple implementation of the idea. We can also envisage models where $h_{\mu\nu}$ in \eqref{SInteraction} is not just $h_{\mu\nu}^{\textrm{MOND}}$ at all scales, but reduces to it only for the physical systems for which MOND is a good description. For example, a very simple way to obtain such a behavior is to define $h_{\mu\nu}$ as
\begin{equation}
h_{\mu\nu}= h_{\mu\nu}^{\textrm{MOND}} \times F(\psi),
\label{genanz}
\end{equation}
where $F(\psi)$ is a phenomenological function of the baryonic matter characterizing the transitions from CDM to MONDian regimes (which, hence, must contain some suitable order parameter, as we shall discuss in Sect.~\ref{Order}).

However, before embarking in a more refined phenomenological study we prefer here to first explore the simplest embodiment of our framework by looking at the case in which $F(\psi)$ is just a constant~\footnote{This framework would correspond to a class of generalized scalar-tensor theories where only the dark matter field is non-minimally coupled to the ``Jordan metric" $\tilde{g}$.}.
In the next section, we will explore the phenomenology of this rather simple model, both considering a high (Sect.~\ref{MoCDM}) and low (Sect.~\ref{MoHDM}) value for the mass $m$ (i.e. cold or hot dark matter) and will finally turn briefly to the more general ansatz \eqref{genanz} in Sect.~\ref{Order}.

\section{Confronting the models with phenomenology}
\label{Section3}
We discuss here the phenomenological predictions of our model of DM-baryon interaction. Since this paper is only an exploratory one, we remain at the qualitative level: more quantitative investigations will be the subject of further studies.

\subsection{MONDian CDM}
\label{MoCDM}
Let us first consider the simplest model where $h_{\mu\nu}= h_{\mu\nu}^{\textrm{MOND}}$ (i.e. $F=1$), and chose a high value for the mass $m$ of the $\chi$ field. 
\paragraph*{Early universe}During the primordial era, the scalar field is expected to evolve rapidly, bringing $\chi$ quickly towards zero, which is the minimum of its potential. We can thus assume that $\partial_t \chi/ \chi$ is large with respect to $m$. It then follows from the definition of $X$ in the appendix that the factor $\chi X$ is extremely small since bounded by $(m_0/m)^{1/2}$, where $m_0=\hbar a_0/c^3 \sim 10^{-69}$~kg is the Compton mass associated to $a_0$. As a consequence of such a negligible $\chi X$, the MONDian effects in the tensor $h_{\mu\nu}$ are cancelled, see Eqs.(\ref{AandB}). Therefore during this era, dark matter decouples from standard matter and the theory reduces to GR+CDM: the model thus reproduces the CMB data.

\paragraph*{Structure formation} The above argument shows that a $\Lambda$CDM's description of the expansion will be valid until late-time cosmology. Therefore the potential wells of dark matter in which baryons collapse to form structures, should develop in a similar way as in the standard model of cosmology. Some differences in the formation process are however expected, since in overdensities the competition between collapse and cosmological expansion may switch on the interaction. In that case, local dynamics would become MONDian for baryonic matter, leading to a bit quicker structure formation than in GR+CDM \cite{Sanders08}. Although the formation of structures in our model is definitely worth studying in details, it goes far beyond the scope of the  present paper since it requires semi-analytical models and numerical simulations.

\paragraph*{Clusters} As we said in the introduction, MOND does explain galactic dynamics but not clusters dynamics, where additional dark matter is needed. This dark matter is essentially needed at the center of clusters, in regions where the acceleration can typically amount to $4 a_0$ \cite{Milgrom:2007cs,Angus:2007mn}. Such a large acceleration implies that, in our model, the effect of the interaction is subdominant so that these regions are indeed filled by some collisionless dark matter. In short, our model would reduce to GR+CDM in the central regions, and MOND in the outer regions. It is a blessing for our model that additional dark matter is needed in MOND only for regions of relatively strong gravity. If one would find an object with gravitational accelerations $g \ll a_0$ still needing some additional DM for MOND to account for it, this would exclude this particular version of our model. 

\paragraph*{Galaxies} We saw that galaxies at the center of rich clusters do also lack some dark matter in MOND. As accelerations are large there, we do expect in our model dark matter in these regions. Low Surface Brightness (LSB) and tidal dwarf galaxies, on the contrary, have internal accelerations everywhere below $a_0$ and must therefore be MONDian, something indeed suggested by recent studies \cite{Milgrom:2007jp, Gentile:2007gp}.

However, just as in the center of clusters where a CDM regime should dominate, the same will happen at the center of high surface brightness (HSB) galaxies. Indeed, for these objects the accelerations in the central regions are also a few times the critical acceleration $a_0$. For instance, the typical acceleration $4a_0$ is attained in the central 3~kpc of the Milky Way. In this particular version of the model we may thus end up with an excess of DM mass in HSB galaxies, although this will depend of course on the precise quantity of mass in such a central mini-halo (which has a small extension of a few kpc). Moreover, the expected central CDM profile is largely unknown in our model since the collapse of astrophysical objects, as we said, will differ from the standard pure CDM theory. Hence, simulations of the structure formation will be needed in order to test this version of the model.

\subsection{MONDian HDM}
\label{MoHDM}

While the HDM scenario has been abandoned long ago by the mainstream research in the cosmological community due to its several problems in explaining structure formation in GR cosmology (hierarchical problem, late formation of structures, etc.), it had, as said, a relevant role in the MOND community due to the fact that some amount of HDM seems to be needed in this scenario. Given that $m$ in our model is a free parameter we briefly explore here how it would behave in the case of small mass.

\paragraph*{Early universe} For what regards cosmology in this HDM scenario, the argument presented in the above section is still valid,  meaning that MONDian effects are negligible until late-times. Hence, the Universe expands according to GR+HDM. It is known that the mass of DM particles does not affect much the angular power-spectrum of the CMB, provided that it reproduces the same total matter density as in the $\Lambda$CDM model. It is even possible to get a similarly good fit with a fermionic particle of $11$~eV as with a $100$~GeV one \cite{Angus:2008qz}.

\paragraph*{Structure formation} As we said, the formation of structure is known to be problematic in a GR+HDM Universe: the computed matter power spectrum is typically failing on scales smaller than $\sim 70$~Mpc. However, the aforementioned competition between expansion and collapse may switch on the interaction leading to a MONDian dynamics in overdensities (below, say, $z=200$ as in the classical MOND+HDM scenario \cite{Angus:2008qz}). It is then known that galactic-scale structures can form before $z=10$, even a bit more quickly than in GR+CDM \cite{Sanders08}. Note that, in this case, virialized objects of galaxy mass are claimed to be the first to form, around $z=10$ \cite{Sanders:1997we}.  Again, this particular model within our general framework will have to be checked in numerical simulations of structure formation, which is beyond the scope of this exploratory paper.

\paragraph*{Galaxies and Clusters} If instead of being very massive, the $\chi$ field is light, it cannot collapse on galactic scales so that the total amount of DM in the center of a HSB is negligible. This solves the problem we have mentioned above, regarding the central regions of these galaxies in the massive case. However dark matter can still collapse on larger scales, and notably in the central region of a cluster, or in a giant elliptical. This way, both clusters and galaxies could be naturally described by such a model.

\subsection{Additional order parameters}
\label{Order}

The two models we just described are essentially characterized by the fact that the field that produces MOND at low accelerations is also the one that behaves as a free massive field at large ones (with respect to $a_0$). The only free parameter remaining is the mass of the field itself. Although these models have already the ability to do better than CDM or MOND alone, they are very simple in the sense that they still rely on the typical value of acceleration to separate in scales, and basically implement a transition from MOND to Newtonian dynamics plus dark matter, instead of the usual transition from MOND to Newtonian dynamics without dark matter. The diversity of gravitational phenomena may however be too rich to be accounted for just with the help of a single order parameter.

We can nonetheless look for a more complicated model that is also able to distinguish between clusters and galaxies, thereby following more closely the idea that dark matter should describe large scales (from cosmology to clusters), whereas MOND should manifest itself only at lower scales. To this end, we write the full tensor $h$ as in Eq.\eqref{genanz}. In this case, the desired phenomenology would arise if $F$ is so to vanish within clusters but saturates to $1$ within spiral galaxies. It is hence necessarily to identify a suitable order parameter. Although other choices may be possible, it is a rather natural idea to use the local density of standard matter to control such a transition \footnote{From the generalized scalar-tensor theory point of view, this would correspond to the most general case in which both dark and baryonic matter couple non-minimally to the metric $\tilde{g}$.}. More precisely, since our equations are local, we should use the density of the {\it medium} of standard matter, i.e. the gas density. Indeed, clusters and galaxies have well-separated {\it medium} densities: $ \rho_{\textrm{clusters}} \ll \rho_{\textrm{galaxies}}$. Consequently we introduce a characteristic scale of density $\rho^*$, with $\rho_{\textrm{clusters}} \ll \rho^* \ll \rho_{\textrm{galaxies}}$, and finally write $h$ as:
\begin{equation}
\label{H2}
h_{\mu\nu}=h_{\mu\nu}^{\textrm{MOND}} \times F(\vert T_\textrm{Mat} \vert/\rho^*).
\end{equation}
The choice of the function $F$ is a matter of model building. For instance $F(x)=x/(1+x^n)^{1/n}$ fits the desired phenomenology. This interpolating function ensures that regions of high gas density will be MONDian (or Newtonian with dark matter depending on the acceleration with respect to $a_0$) whereas in low density regions the DM field decouples from matter and the models boils down to CDM. 

The exact quantification of $\rho^*$ requires some considerations about astrophysics. As we saw, spiral galaxies are well described by MOND, meaning that $\rho^*$ must be smaller than the typical density of the interstellar HI gas filling spiral galaxies. Thus $\rho^*$ should be less than $5\times 10^{-22} kg/m^{3}$.

In this approach, we want clusters to be fully described by CDM, meaning that the DM field should decouple from baryons in almost all their volume. Even if the DM field couples to baryons in some small regions of the cluster (within spiral galaxies belonging to the cluster for instance), the overall dynamics will satisfactorily be described by the CDM scheme up to minor differences. The typical value of the intracluster hot gas density can e.g. be deduced from data found in Table 1 of \cite{Angus:2007mn}, and we get $\rho_{\textrm{Intraclusters}} \sim 10^{-25} kg/m^{3}$.

Then we conclude $\rho^*$ must lie in the following range (in $kg/m^{3}$): $10^{-25} \alt \rho^* \alt 5 \times 10^{-22}$. Having explicited all the features of the model, we now proceed to detailed phenomenology and qualitative predictions.

\paragraph*{Early universe, structure formation and clusters} Following the reasoning of the previous sections it should be clear that cosmology at early times still reduces to $(\Lambda)$CDM given that $h_{\mu\nu}^{\textrm{MOND}}$ is small. So, both the CMB spectrum and the structure formation are explained by the model, provided that we chose a large mass for $\chi$. In fact, because the density of cosmological matter decreases with time due to the expansion, cosmology at late-time is also $\Lambda$CDM in this model. Finally, clusters are by construction in a CDM regime.

\paragraph*{Galaxies} In this model elliptical galaxies must also be in a CDM regime because, as it is well-known, they have very low local gas density. In fact, the status of these galaxies is unclear from an observational point of view. For instance, strong lensing does indicate a cuspy core as in CDM theory \cite{Chen:2008xpa}. Moreover, there are several strong lensing images around ellipticals which cannot be fitted by relativistic theories of MOND unless some more DM is added \cite{Shan:2008sn}. On the other hand, both weak lensing and dynamics are in favor of an isothermal profile \`a-la-MOND \cite{Gavazzi:2007vw, Tiret:2007kq}. More observational evidences are thus needed to discriminate between CDM and MOND in these systems(see e.g. \cite{Tian:2008sg}). 

Tidal dwarfs galaxies, as well as LSB and HSB spirals are well described by MOND if almost no dark matter is present in the central regions (e.g., in the Milky Way, in a sphere of 10 kpc). The central mini-halo expected in our model, and discussed in the previous sections, may be problematic in this regard. Furthermore, the local density quickly decreases in the direction orthogonal to the galactic plane and rapidly approaches typical intracluster densities. These two issues together imply that in this model one should expect a galactic dynamics which is not purely MONDian but rather may have significant dark matter contributions. Lacking a detailed simulation of the galaxy formation in this framework it is however unclear how much the model will deviate from the standard predictions of MOND\footnote{A similar issue appears in another approach of the literature \cite{Blanchet:2008fj} in which DM particles carrying also a four-vector are considered. In this model, the energy density involves a monopolar and a dipolar term: the monopolar term can play the role of CDM in the early Universe, whereas the dipolar term, under certain circumstances, creates the MOND regime through what can be called ``gravitational polarization". However, the monopolar part is still weighting in that case. Therefore, the DM particles should not collapse too much at galactic scales in order to avoid an excess mass problem. Authors of Ref.\cite{Blanchet:2008fj} then evoked a weak clustering hypothesis to make their model consistent with the data.} (which moreover can depend on the precise value of the scale $\rho^*$ which can be somehow fine-tuned). 

With regards to weak lensing around galaxies similar uncertainties apply. However, given that the theory is built to reduce to $\Lambda$CDM at large scales, weak lensing around galaxies will be induced by the metric $g$, and not by the MONDian one $g+h$. Then, some deviations can be expected from the standard predictions of  $\Lambda$CDM due to the fact that this model will require to have somewhat less dark matter inside the galaxy.

As a final comment, it is an interesting feature of the model that fluffy globular clusters in the outskirst of galaxies may be either MONDian or Newtonian depending on their gas density. Gas poor dwarf spheroidal satellites of the Milky Way may also behave according to CDM rather than to MOND. Consequently, these systems represent a severe test to disentangle different implementations of our general framework, and are thus worth studying in more detail (see e.g. \cite{Baumgardt:2005ie}).  

\section{Conclusion}
\label{Section4}
In this paper we explored some possible ways to reconcile theories of MOND and CDM, in order to merge their successes in a single framework. To this end, we introduced in Sect.~\ref{Section2B} a suitably chosen interaction between dark matter, baryons and gravity, such that CDM and MOND appear in different physical regimes of the same theory. We also showed that this framework may have a natural interpretation within a class of generalized scalar-tensor theories which couple non-minimally dark matter (and possibly baryons) to the metric.  

Although we provided only a qualitative analysis, the simplest models of this kind already have the ability to achieve a nice unification of MOND and dark matter scenarios (cold or hot dark matter), as we have shown in Sects.~\ref{MoCDM} and \ref{MoHDM}. Moreover these models, besides the unification of CDM and MOND, also generically predict new phenomena that may falsify them. In particular, a quantitative study of structure formation and hence an estimate of the distribution of dark matter within structures are natural tests of the framework. This is left for future investigations.
 
We have also discussed the possibility that this class of models may have to incorporate extra order parameters besides the Milgrom's acceleration scale $a_0$. We showed that our framework is general enough to handle such a refined picture, for instance by generalizing the simple ansatz for $h$ we made in Eq.\eqref{hMond} into the one given in Eq.\eqref{genanz}. More specifically, we considered in Sect.~\ref{Order} a model involving a critical medium density separating the MOND regime from the CDM one, so that in spiral galaxies it would be MONDian whereas in clusters and elliptical galaxies it would be Newtonian with CDM.
 
There are also some theoretical issues to address, like the particle physics origin of our peculiar coupling between baryons and dark matter. In this sense its interpretation within an extended scalar-tensor theory could be a useful guide. Furthermore it remains to be investigated whether this class of models may suffer from consistency problems, like the stability issues discussed in  Ref.\cite{Bruneton:2007si}. Although these issues deserve a careful treatment, the situation may differ from the ones already considered in the literature because of the freedom we have in generalizing the ansatz for $h$, which could be used to cure these stability problems without going beyond our general scheme. 

Let us note that the idea to introduce the medium density as another order parameter does not depend on the particular framework considered in this paper. This hypothesis has promising and intriguing consequences: fluffy globular clusters in the outskirst of galaxies may be either MONDian or Newtonian depending on their gas density, while gas poor dwarf spheroidal satellites of the Milky Way would behave according to CDM rather than to MOND. The fact that these systems might indeed be prolematic for MOND \cite{Zhao:2005xk,Angus:2008vs} makes this hypothesis in itself worth investigating in detail in the future.

As a concluding remark, we stress that since this framework is able to generate various models, it can offer different points of view on the mass discrepancy problem. As a consequence, it is our hope to extract from it some model-independent statements about how dark matter and MOND could coexist in a single description accounting for gravitational physics at all scales. 

\begin{acknowledgments}
The authors wish to thank L. Danese, G. Esposito-Far\`ese, V. Faraoni and P. Salucci for constructive suggestions and comments, and F. Girelli for his help on early stages of the work. BF acknowledges financial support from the Belgian FNRS.
\end{acknowledgments}

\appendix
\section*{Appendix}
Let us briefly comment on the construction of the tensor $h$, defined in Eq.\eqref{hMond} and Eq.\eqref{AandB}. It follows the reasoning found in \cite{Bruneton:2007si}: As the $\chi$ field propagates freely in vacuum, it must assume the following form around a spherical body of mass $M$:
\begin{equation}
\label{behavchi}
\chi \approx - \alpha \frac{G M}{r c^2} e^{- m r},
\end{equation}
where $m$ is the mass of the $\chi$ field and $\alpha$ is the strength of the coupling with matter. This is the standard result for massive scalar-tensor theories which must hold here to the leading order for the reason that, in practice, astrophysical bodies that generate $\chi$ are compact enough to be in a Newtonian regime\footnote{Taking into account the dilute baryonic gas that fills galaxies is however a subtle issue, see \cite{Bruneton:2007si}. For what follows we assume Eq.\eqref{behavchi} to be valid, even though solving the whole problem may eventually lead to a revised definition of $h$.} (e.g. stars). Then $M$ and $r$ can be expressed by combinations of local quantities. Let us define 
\begin{eqnarray}
\varpi &=& \left(\frac{\sqrt{(\partial \chi)^2}}{\chi}-m\right)^{-1}, \nonumber \\
X&=& \frac{\sqrt{\alpha a_0}}{c} \sqrt{\frac{\varpi}{- \chi e^{m \varpi}}},
\end{eqnarray}
that respectively behave as $r$ and $( a_0 r^2/ GM)^{1/2}$ when $\chi$ assumes the previous form. Then the tensor $h$ defined in the text is construct so as to give an effective metric $g+h$ that is of the MONDian type:
\begin{eqnarray}
\tilde g_{\mu\nu} &=& g_{\mu\nu}+h_{\mu\nu}\nonumber \\
&=&\left[1- \frac{2\alpha^2GM}{rc^2}e^{-m r}
+\frac{2\sqrt{GMa_0}}{c^2}\,
u\left(\sqrt{\frac{a_0 r^2}{GM}}\right)\right] g_{\mu\nu} \nonumber \\
&+&\frac{4\sqrt{GMa_0}}{c^2} \delta_\mu^r \delta_\nu^r
+ \mathcal{O}\left(\frac{1}{c^4}\right),
\label{gtildeM}
\end{eqnarray}
at the leading post-Newtonian order. The function $u$ appearing in Eq.\eqref{AandB} ensures the transition from MOND to Newton as a function of the strength of the gravitational force with respect to $a_0$, and can be taken to be $u(x)= (1+x)^{-1} + \ln(1+x)$. 
Note that the quantity $(\partial \chi)^2$ could be negative in non-static situations, so that, in the definition of $\varpi$, it could be convenient to replace it by $\vert(\partial \chi)^2\vert$.

\bibliographystyle{apsrev}
\bibliography{BiblioMOND}

\end{document}